# A Novel Lie Algebra of the Genetic Code over the Galois Field of Four DNA Bases.

Robersy Sánchez[1, 2], Ricardo Grau[2]

[1] Research Institute of Tropical Roots, Tuber Crops and Banana (INIVIT). Biotechnology group. Santo Domingo. Villa Clara. Cuba.
E-mail: robersy@uclv.edu.cu

[2] Center of Studies on Informatics. Central University of Las Villas. Villa Clara. Cuba. E-mail: rgrau@uclv.edu.cu

Corresponding author: Robersy Sánchez
Mailing address: Apartado Postal 697. Santa Clara 1, CP 50100. Villa Clara, Cuba.

Emails: robersy@uclv.edu.cu
robersy@inivit.co.cu

**Abstract**.

By starting from the four DNA bases order in the Boolean lattice, a novel Lie Algebra of the genetic code is proposed. Here, the principal partitions of the genetic code table were obtained as equivalent classes of quotient subspaces of the genetic code vector space over the Galois field of the four DNA bases. The new algebraic structure shows strong connections among algebraic relationships, codon assignment and physicochemical properties of amino acids. Moreover, a distance function defined between codons in the Lie algebra was demonstrated to have a linear behavior respect to physical variables such as the mean of amino acids energy in proteins. It was also noticed that the distance between wild type and mutant codons approach smaller values in mutational variants of four genes, i.e, human phenylalanine hydroxylase, human β-globin, HIV-1 protease and HIV-1 reverse transcriptase. These results strongly suggest that deterministic rules in genetic code origin must be involved.

*Key words or phrases*: Genetic code vector space – Genetic code algebra – Genetic code Lie algebra – gene mutation.

# 1. Introduction

The nature of genetic code is now fairly well known. From the second half of 20th century, many attempts have been made, just to understand its internal regularity (Bashford and Jarvis, 2000; Bashford and Tsohantjis, 1998; Beland and Allen, 1994; Crick, 1964; Eck, 1963; Epstein, 1966; Jiménez-Montaño, 1966; Jukes, 1977; Volkenshtein, 1985). The Code represents an extension of the four-letter alphabet of deoxyribonucleic (DNA) bases: Adenine (A), Guanine (G), Cytosine (C), and Thymine (T) or Uracil (U) in ribonucleic acid (RNA). As established, chemical pairing by hydrogen bonds occurs between G≡C and A=T (U), which means G is complementary base of C and A to T (U) or *viceversa*.

Furthermore, an association between codons having U at second base position and hydrophobicity of the amino acids was also observed, i.e. for amino acids I, L, M, F, V (one-letter symbol of amino acids). Whereas those codons having A at second base position code to hydrophilic or polar amino acids, i.e., D, E, H, N, K, Q, Y (Crick, 1968). Epstein (1966) has stated that amino acids cannot be randomly allocated by just considering the features of the genetic code -fully discussed by Crick (1968)- and particularly we believe that the order of codons must reflect their physicochemical properties. Anyway, an optimal distribution of the table must be assumed. Gillis et al. have suggested that genetic code can be optimised by limiting translation errors (Gillis et al, 2001), so that an optimal codon order should be established. On the other hand, a deterministic connection of codon assignment with physicochemical properties of amino acids can be stated (Lehmann, 2000; Robin et al. 1999).

Recently it has been published the Boolean algebra of the four DNA bases and Boolean algebra of the genetic code (Sánchez et al., 2004a, 2004b and 2005a). The Boolean structures were defined using the most elementary physicochemical properties of bases, the hydrogen bond numberings and base chemical types (both purines or pyrimidines). Now, from the Boolean lattice of four DNA bases, a vector space and a Lie algebra is proposed, using a new algebraic standpoint of the genetic code. The aim of present work is connecting the algebraic relationships among codons to physicochemical properties of amino acids, codon assignment and single point mutation in genes.

**2. Theoretical Model**

The genetic code vector space is presented over the Galois field of four bases ($GF(4)$). The starting point is the bijection $f$: {A, C, G, U} $\to$ {($a_i$, $a_{i+1}$)} from the base set to the set of binary duplets ($a_i$, $a_{i+1}$), where $f(X) = (a_i, a_{i+1})$ and $a_i \in \{0, 1\}$. The biological criterion to state a bijection between the base set and the binary duplet representation of $GF(4)$ elements is the base complementarity in DNA duplex molecule. That is, for instance, for the set {A, C, G, U} we have the equalities: $f(A) = (0, 0)$, $f(C) = (1, 0)$, $f(G) = (0, 1)$, $f(U) = (1, 1)$. As a result the sum of binary digits corresponding to DNA complementary bases is always the duplet (1, 1). On this criterion we have eight ordered base sets: {G, A, U, C}; {G, U, A, C}; {C, A, U, G}; {C, U, A, G}; {A, C, G, U}; {A, G, C, U}; {U, C, G, A} and {U, G, C, A}. Any of these base orders leads to same algebraic properties of a genetic code vector space over $GF(4)$. To start our study we have chosen the base orders {G, U, A, C} and {A, C, G, U} which have been derived from previous genetic code algebraic structures (Sanchez et al., 2004a, 2004b, 2005a and 2005b).

*2.1. Genetic code vector space*

From the Boolean lattices of four DNA bases, the primal order {G, U, A, C} was considered (Sanchez et al., 2004a, 2004b and 2005a). Next, the algebraic operations, sum ("+") and product ("•") in the set of bases were introduced under the classical field definition of a Galois field structure (Kostrinkin, 1980). If a Galois field structure in the ordered base set

{G, U, A, C} is assumed, then, a bijective correspondence with the Galois field of four elements ($GF(4)$) determines a unique definition of sum and product of bases.

Particularly, an isomorphism with the Galois field is defined in the set $(Z_2)^2=\{(0, 0), (1, 0), (0, 1), (1, 1)\}$ ($Z_2=\{0, 1\}$), i.e. a unique $GF(4)$ up to isomorphism exists, so that the isomorphism: G↔(0, 0), U↔(1, 0), A ↔(1, 0), C ↔(1, 1) is stated. An isomorphism between the primal Boolean lattice of the four bases and the Boolean lattice $((Z_2)^2, \vee, \wedge)$ was previously reported and led to a similar binary representation (Sánchez et al., 2004a). The operation tables of the Galois field on the set $B$ of the four DNA bases are shown in Table 1.

TABLE 1

The set of four bases $B$ with the operation "+" is evidently an Abelian group, denoted by $(B, +)$. Next, an Abelian group on the set of codons $C_g$ will be the direct third power of the group $(B, +)$, i.e.:

$$(C_g, +) = (B, +) \times (B, +) \times (B, +)$$

Taking into account the uniqueness of the $GF(4)$ up to isomorphism, the element $\alpha \bullet x \in (B, +)$ may be later defined, for all elements $\alpha \in GF(4)$ and for all elements $x \in (B, +)$. Moreover, it can be noticed that the group $(B, +)$ have a vector space structure over $GF(4)$. Similarly, it is obtained the three-dimensional vector space $VG$ of the group $(C_g, +)$ over $GF(4)$.

In vector space $VG$ the product of two vectors $X = X_1X_2X_3 = (X_1, X_2, X_3)$ and $Y=Y_1Y_2Y_3 = (Y_1, Y_2, Y_3)$ ($X, Y \in (C_g, +)$) will be the product of each component of them, so that:

$$X_1X_2X_3 \bullet Y_1Y_2Y_3 = (X_1 \bullet Y_1, X_2 \bullet Y_2, X_3 \bullet Y_3)$$

Finally, this structure may be extended to *N*-dimensional sequence spaces (*S*) consisting of the set of all $64^N$ DNA sequences having *N* codons. Evidently, there is a bijection between this set and the set of all *N*-tuples $(x_1,...,x_N)$ where $x_i \in (C_g, +)$. Then, the algebraic structure of $(C_g, +)$ can be extended to the set *S* represented by all *N*-tuples $(x_1,...,x_N) \in ((C_g, +))^N$. To be precise, the set *S* is represented by the direct sum ($\oplus$) of the *N* groups $(C_g, +)$:

$$(S, +) = (C_g, +) \oplus (C_g, +) \oplus ... \oplus (C_g, +) \quad (N \text{ times})$$

The direct sum $(S, +)$ becomes the set of all sequences $g = (c_1, c_2,..., c_N)$, where $c_i \in (C_g, +)$. Likewise, the *N*-dimensional vector space $(VG)^N$ is obtained. We shall define the product of two sequences $g_1$ and $g_2$ by the components (bases):

$$X_1 X_2 ... X_N \bullet Y_1 Y_2 ... Y_N = (X_1 \bullet Y_1,\ X_2 \bullet Y_2,\ ...,\ X_N \bullet Y_N)$$

*2.2. Lie algebra of the genetic code.*

In the vector space of the genetic code over *GF*(4) a commutator can be defined, as in the vector product in the three-dimensional vector space on $R^3$. So, analogous to the classical vector product for every pair of codons ($X = X_1 X_2 X_3 = (X_1, X_2, X_3)$ and $Y = Y_1 Y_2 Y_3 = (Y_1, Y_2, Y_3)$ ($X, Y \in (C_g, +)$), the codon $[X, Y]$ is defined:

$$[X, Y] = (X_3 \bullet Y_2 + X_2 \bullet Y_3,\ X_3 \bullet Y_1 + X_1 \bullet Y_3,\ X_2 \bullet Y_1 + X_1 \bullet Y_2)$$

It can proved that for all $\alpha, \beta \in GF(4)$ and for all $X, Y \in (C_g, +)$ the commutator must satisfy the following properties:

i. $[\alpha \bullet X + \beta \bullet Y, Z] = \alpha \bullet [X, Z] + \beta \bullet [Y, Z]$

ii. $[X, Y] + [Y, X] = (G, G, G) = GGG$

iii.  $[X, [Y, Z]] + [Y, [Z, X]] + [Z, [X, Y]] = $ GGG    (Jacobi identity)

If a vector space on a field $F$ with a commutator satisfies the above properties, then, it is called Lie algebra over that field $F$. Since the characteristic of the field $GF(4)$ is 2, unlike to the vector product in the three-dimensional vector space on $R^3$, it follows the symmetric property:

$$[X, Y] = [Y, X]$$

Next, commutation between two sequences, $g_1 = (c_{11}, c_{12}, \ldots, c_{1N})$ and $g_2 = (c_{21}, c_{22}, \ldots, c_{2N})$, ($g_1, g_2 \in (VG)^N$) will be defined by components. That is, the commutation $[g_1, g_2]$ is the vector:

$$[g_1, g_2] = ([c_{11}, c_{21}], [c_{12}, c_{22}], \ldots, [c_{1N}, c_{2N}])$$

The $N$-dimensional vector space $(VG)^N$ having this commutation operation is a $N$-dimensional Lie algebra $(LG)^N$ (Pontriaguin, 1978), i.e. the $N$-dimensional Lie algebra $(LG)^N$ will be $N$ times the direct sum of $LG$.

## 3. Results and Discussion

Both vector space and Lie algebra must allow understanding the logic involved in genetic code. Neutral elements of product and sum operations of primal $LG$ code for glycine, the simplest amino acid. Amino acids differ essentially in side chains or $R$ groups, which are bound to α-carbon of the backbone. For instance, glycine has the simplest $R$ group, a hydrogen atom. Thus, from a molecular standpoint, glycine structure is present in all amino acids, that is, glycine posses basically the structure from which every amino acid is built. As a result, the $LG$ built using the base order {G, U, A, C} should be considered a biologically significant order between the eight possible ordered.

*3.1. Genetic code and codon assignment*

Genetic code has been usually presented as a four-column table where codons are placed following the second base order (Table 2). Codons orderings or base placements in codons may change elsewhere, so that hydrophobic or hydrophilic coded amino acids may also be found in different columns in that table.

By contrast, the present proposal involves the basic entries of the genetic code tables pointing out relationships between vector subspaces, codon assignments and physicochemical properties of amino acids.

Table 2

In the vector space of genetic code, the vector subspace $C_G$ on the subset of codons $\{X_1GX_2\}$ is a normal subspace. The quotient subspace $VG/C_G$ of the genetic code vector space is formed by the elements:

$$\{C_G, X_U + C_G, X_A + C_G, X_C + C_G\}$$

Where, $X_U$, $X_A$ and $X_C$ are arbitrary elements of the codon subsets: $\{X_1UX_2\}$, $\{X_1AX_2\}$ and $\{X_1CX_2\}$. For instance, codons belonging to set $\{X_1UX_2\}$ may be represented by any sum,

$$\{X_1UX_2\} = \text{AUG} + C_G = \text{AUG} + \{X_1GX_2\} \text{ or } \{X_1UX_2\} = \text{CUA} + C_G = \text{CUA} + \{X_1GX_2\}$$

In particular, these means

$$\text{AUC} = \text{AUG} + \text{GGC or AUC} = \text{CUA} + \text{UGU}$$

where $\text{AUG} \in \{X_1UX_2\}$, $\text{CUA} \in \{X_1UX_2\}$, $\text{GGC} \in \{X_1GX_2\}$ and $\text{UGU} \in \{X_1GX_2\}$

The quotient subspace $VG/C_G$ is a partition of the set of codons in 4 equivalence classes. Every class has the same number of elements $C_G$, i.e. 16 codons, which correspond to the 4 main columns of Table 2. Subsets $X_T + C_G = \{X_1UX_2\}$, $X_A + C_G = \{X_1AX_2\}$ and $X_C + C_G =$

{$X_1CX_2$} are cosets of the linear subspace $C_G$ and consequently, they are affine subspaces of the vector space *VG*.

The neutral vector subspace $C_{GG}$ on the subset of codons {*GGX*} is also a normal subspace. Likewise, the quotient vector space $VG/C_{GG}$ of the code vector space can be defined. As a result there are 16 equivalent classes of codons, having every class 4 codons with the first two bases constant. Such an arrangement can be noticed in the standard table (see Table 2). We have, for instance,

$$AUG + C_{GG} = \{AUG, AUU, AUA, AUC\}$$

$$CAG + C_{GG} = \{CAG, CAU, CAA, CAC\}$$

for two of these classes (see Table 2).

As pointed out before by Crick, the first two bases of codon determine the physicochemical properties of amino acids (Crick, 1968). The 4 coded amino acids of every class are either the same or show very similar physicochemical properties. From the above considerations, the algebraic relationships among codons can be related to physicochemical similarities of coded amino acids.

Finally, it has been found 3 normal subgroups $C_{GGT}$, $C_{GGA}$ and $C_{GGC}$ on the subsets {GGG, GGU}, {GGG, GGA} and {GGG, GGC} respectively. The corresponding quotients subgroups $C_g/C_{GGT}$, $C_g/C_{GGA}$, $C_g/C_{GGC}$ can be described, having the subgroup $C_g/C_{GGA}$ a remarkable biological meaning. In such a case, elements are subset of two codons, each one of them showing the same type of base, i.e. purine (*R*, $R \in$ {G, A}) or pyrimidine (*Y*, $Y \in$ {U, C})(see Table 3). That is, for every class of the quotient subgroup their base triplets have the form: $X_1X_2R$ or $X_1X_2Y$ ($X_1$, $X_2 \in$ {G, T, A, C}). Codons in each class code either for the same amino acid or two different amino acids with highly similar physicochemical properties (Table 2 and 3). Amino acids, from just two synonymous codons have both codons in the same class. Therefore the quotient subgroup preserves all chemical types of bases in codons.

Table 3

According to these results, we have described the principal partitions of the genetic code table, a discovery from the second half of past century (Crick, 1968; Epstein, 1966). These results also suggest that there is a strong connection between codon algebraic properties, codon assignment and physicochemical properties of amino acids in such a way that the origin of genetic code seems not to be at random. It is amazing that any base order leads to similar genetic code partitions, however, for each particular base order it is possible bring out particular biological remarks.

*3.2. Genetic code Lie algebras and physicochemical properties of amino acids.*

If two codons $X = X_1X_2X_3$ and $Y = Y_1Y_2Y_3$ have their commutator equal to GGG (analogous to vector in $R^3$) then:

$$[X, Y] = [X, X + Y] = [Y, X + Y] = GGG \qquad (2)$$

After the distributive law and being $[X, X] = GGG$ for all $X \in LG$, then the last equalities are evident. In particular, for every $X_1UX_2$ there is a codon $Y_1AY_2 \neq NNN$ such that:

$$[X_1UX_2, Y_1AY_2] = [X_1UX_2, X_1UX_2 + Y_1AY_2] = [Y_1AY_2, X_1UX_2 + Y_1AY_2] = GGG \qquad (3)$$

By analogy to the 3-dimensional vector space, the codons that satisfy the property $[X, Y] = GGG$ are called parallel codons. As a result the set of codon is sorted into 21 subsets of parallel codons $\{X, Y, X + Y\}$ (see Table 4). Moreover, if codons $X = X_1X_2X_3$, $Y = Y_1Y_2Y_3$ and $Z = Z_1Z_2 Z_3$ belong to a subset of parallels codons then, $X = \lambda \bullet Y$ and $Y = \lambda \bullet Z$, where $\lambda \in \{U, A, C\}$ and the determinant:

$$\begin{vmatrix} X_1 & X_2 & X_3 \\ Y_1 & Y_2 & Y_3 \\ Z_1 & Z_2 & Z_3 \end{vmatrix} = G \quad \text{(Or (0, 0) in the binary representation)}$$

Explicitly, by analogy to three-dimensional vector space, we could say that codons $X$, $Y$ and $Z$ are collinear. So, for every pair of collinear codons $X_1UX_2$ and $Y_1AY_2$, coding for non-polar and polar amino acids respectively, there is a parallel codon $Z_1CZ_2$ that codes for amino acids of middle polarity, according to Grantham polarity scale (1974).

Table 4

Next, 21-groups {GGG, $X$, $Y$, $X + Y$} are formed by adding GGG to each subset of parallel codons and every of them is a Klein four group. At the same time, they are vector subspaces and Lie subalgebras of *LG*. For instance, the group {GGG, GGU, GGA, GGC} becomes the neutral subgroup $C_{GG}$. Except for the neutral subgroup of glycine codons, the maximal absolute differences in Δ-free energy values of transfer from water to octanol among amino acids coded by parallel codons are, in general, extreme (Table 4). Of course, maximal differences are established among codons $X_1UX_2$ (coding for hydrophilic amino acids) and $Y_1AY_2$ (coding for hydrophobic amino acids).

Provided that operations sum and difference are equivalent, for the parallel subgroup, it is hold the following symmetrical relationships:

$$X_1UX_2 \pm Y_1AY_2 = Z_1CZ_2, \qquad (4)$$

This symmetrical relationship is linked to the physicochemical properties of amino acids. For amino acid pairs coded by parallel codons $X_1UX_2$ and $Y_1AY_2$, the mean of its polarities is, in general, near to the polarities of amino acids coded by codons $X_1CX_2$ (see Table 4). The symmetrical relationships (4) reflect the symmetrical space distribution of amino acids in protein structure. Hydrophobic regions in proteins –coded by codons $X_1UX_2$– are surrounded by hydrophilic regions –coded by codons $X_1AX_2$– so that hydrophilic amino acids tend to buffer the effect produced by hydrophobic amino acids (Volkenshtein, 1983). As a result, the protein molecule becomes of middle polarity. So, the algebraic relationship between codons in *LG* reflexes the physicochemical interactions among their corresponding amino acids in proteins.

Furthermore, it must be expected that most of the single point mutations in proteins should not involve parallel codons. For example, point mutations called transversions which alter the chemical type of bases are frequently more dangerous than transitions, which preserve this property. In fact, transversions at first and second base positions of codons are the most dangerous mutations. Thus, transitions are more frequent in nature (Yang, 2000). Going from one codon to any parallel one at least one transversion is needed (Table 4). In general, changes between parallel codons involve two transversions at first and second base.

*3.2. Linear transformations in VG and $(VG)^N$.*

Lineal transformations will allows us to study the mutational pathway in the *N*-dimensional space $(VG)^N$ of DNA sequences. So, our starting point will be the linear transformation on *VG*. The algebraic operations over the base triplet are equivalent to the derivation of new codons by means of base substitution mutations in the ancestor codon. In particular, we are interested in the automorphisms. Given that these transformations are invertible, the mutation reversions are forecasted. In addition, it is well known that the set *G* of all automorphisms is a group. Notice that the endomorphism *f* will be an automorphism if, and only if, the determinant of its automorphism representing matrix is equal to G.

As we said above vector space *VG* can be sorted into 21 subset of parallel codons, each one of them form a Klein four group by adding codon GGG (Table 3). The set of 21 Klein four groups is closed to commutator operation, i.e. if $K_1$ and $K_2$ are Klein four groups then, for any codons $X_1X_2X_3 \in K_1$ and $Y_1Y_2Y_3 \in K_2$ there is the codon $Z_1Z_2Z_3 = [X_1X_2X_3, Y_1Y_2Y_3] \in K_3$, where $K_3$ is a Klein four group of codons. This is a consequence of the fact:

$$[X_1X_2X_3, Y_1Y_2Y_3] = \lambda \bullet [X'_1UX'_3, Y'_1UY'_3] = Z_1Z_2Z_3$$

, where $X_1X_2X_3, X'_1UX'_3 \in K_1$; $Y_1Y_2Y_3, Y'_1AY'_3 \in K_2$; $Z_1Z_2Z_3 \in K_3$ and $\lambda \in \{U, A, C\}$.

As a result, any endomorphism on *VG* transforms one Klein four group into another. If one element of a Klein four group $K_1$ is transformed –by means of an endomorphism on *VG*– into one element of a Klein four group $K_2$ then the rest of elements of $K_1$ are transformed into the corresponding elements of $K_2$.

Now, let $S_k$ be the subset of codons conserving the same base position $k \in \{1, 2, 3\}$. Then, according to the group theory (Kostrinkin, 1980), the set $St(k)$ of automorphisms $f \in G$ that preserves the base position $k$ is a subgroup of $G$, that is:

$$St(k) = \{f \in G, \text{ such that: } f(X_1X_2X_3) \in S_k\} \subset G$$

This subgroup could be called the stabilizer subgroup of the group *G* that fix base position *k*. Next, we take into consideration that most frequent mutations observed in codons preserve the second and the first base position, Accepted mutations on the third base are more frequent than on the first base, and, in turn, these are more frequent than errors on the second base (Friedman and Weinstein, 1964; Woese, 1965; Parker, 1989). These positions, however, are too conservatives with respect to changes in polarity of the coded amino acids (Alf-Steinberger, 1969). Consequently, the effects of mutations are reduced in the genes and the accepted mutations decreased from the third base to the second. So, we have to expect that most frequent automorphisms should preserve the second and the first base positions, i.e. most frequent mutation should belong to the subgroups $St(2)$ and $St(1)$. In particular, the automorphism subgroup $St(2)$ map an affine subspace of the quotient subspace $VG/C_G$ into itself, while the subgroup $St(1)$ map an affine subspace of the quotient subspace $VG/C_{GG}$ into itself.

As a result, automorphisms $f \in St(2)$ and $g \in St(1)$ transform each element of a Klein four group $K_1$ into the corresponding element of a Klein four group $K_2$, i.e. the first codon of the parallel subgroup $K_1$ into the first codon of the parallel subgroup $K_2$ and so on. For instance, codons AUC, CAU and UCA are transformed into codons UUA, AAC and CCU respectively, by means of automorphism $\begin{pmatrix} A & U & U \\ G & A & A \\ C & A & C \end{pmatrix}$. In general we have:

$$X_1 X_2 X_3 \begin{pmatrix} a_{11} & a_{12} & a_{13} \\ a_{21} & a_{22} & a_{23} \\ a_{31} & a_{32} & a_{33} \end{pmatrix} = \lambda \bullet X_1 U X_3 \begin{pmatrix} a_{11} & a_{12} & a_{13} \\ a_{21} & a_{22} & a_{23} \\ a_{31} & a_{32} & a_{33} \end{pmatrix} = \lambda \bullet Y_1 U Y_3 = Y_1 Y_2 Y_3$$

where $\lambda \in \{G, U, A, C\}$; $X_1X_2X_3$, $X_1UX_3 \in K_1$ and $Y_1Y_2Y_3$, $Y_1UY_3 \in K_2$.

Now we can analyze endomorphisms on the *N*-dimensional vector space $(VG)^N$ of DNA sequences. The endomorphism ring $\text{End}((VG)^N)$ is isomorphic to the ring of all matrices $(A_{ij})$, where $A_{ij} \in \text{Hom}(VG_i, VG_i)$ ($i = 1, .., N$), with the traditional matrix operations of sum and product. In our particular case, the principal diagonal element are matrices $A_{ii} \in \text{End}(VG)$ (or $A_{ii} \in \text{Aut}(VG)$) and non-diagonal element are null-matrices. Mutations in DNA sequences will correspond to automorphisms when $A_{ii} \in \text{Aut}(G_i)$ for all triplets in the DNA sequence.

As has been pointed out in a previous paper (Sanchez et al. 2005b), since evolution could not happen without the genetic recombination, it is biologically relevant that the homologous recombination that involves a reciprocal exchange of DNA sequences −e.g. between two chromosomes that carry the same genetic loci− algebraically corresponds to the action of two automorphism pairs (see Fig. 1). For instance, the pair $f$ and $f^{-1}$ acts over the homologous strands α and β to turn out the homologous reciprocal recombinants $f(\alpha)$ and $f^{-1}(\beta)$, and the pair $g$ and $g^{-1}$ acts over the homologous strands α' and β' to turn out the homologous reciprocal recombinants $g(\alpha')$ and $g^{-1}(\beta')$. As a result two reciprocal recombinant DNA sequences are generated. Next, due to the symmetrical property of the commutator we have −for homologous recombination− the following identities:

$[\alpha, \beta] = [f(\alpha), f^{-1}(\beta)]$

$[α', β'] = [g (α'), g^{-1} (β')]$

$[α, f(α)] = [β, f^{-1} (β)]$

$[α', g (α')] = [β', g^{-1} (β')]$

That is, the symmetrical property of commutator reflects the balanced distribution of the genetic information in homologous recombination preserving commutator values between DNA homologous pairs. If it were not possible to exchange material between (homologous) chromosomes, the content of each individual chromosome would be irretrievably fixed in its particular alleles. When mutations occurred, it would not be possible to separate favourable and unfavourable changes (Lewin, 2004). Hence, the study of the automorphism subgroup involved in this transformation –the homologous recombination– could disclose new rules of molecular evolution process unknown so far.

*3.4. Genetic code partition with base order {A, C, G, U}*

Analogous to the last sections we can introduce partitions of the genetic code vector space over *GF*(4) when it is used the base order {A, C, G, U}. This base order allows us introduce a distance function.

The starting point is the bijection $f$: {A, C, G, U} → {$(a_i, a_{i+1})$} from the base set to the set of binary duplets $(a_i, a_{i+1})$, such that $f$ (A) = (0, 0), $f$ (C) = (1, 0), $f$ (G) = (0, 1) and $f$ (C) = (1, 1). Next, taking into account the biological importance of base position in the codons we define de bijection φ:

$$φ (X_1X_2X_3) = (f(X_3), f(X_1) f(X_2)) = (a_0, a_1, a_2, a_3, a_4, a_5)$$

, where $f(X_3) = (a_0, a_1), f(X_1) = (a_2, a_3)$, and $f(X_2) = (a_4, a_5)$ and $a_i \in \{0, 1\}$.

Then the distance function is defined as:

$$d(X_1X_2X_3, Y_1Y_2Y_3) = d((x_0, x_1, x_2, x_3, x_4, x_5),(y_0, y_1, y_2, y_3, y_4, y_5)) = \sum_{i=0}^{5} \frac{\delta_i}{2^{5-i}}$$

, where $\delta_i = \begin{cases} 1 & \text{if } x_i = y_i \\ 0 & \text{others} \end{cases}$

Now a pseudo-norm can be defined as the distance between codons AAA and $X_1X_2X_3$. That is:

$$d(AAA, X_1X_2X_3) = \|X_1X_2X_3\|$$

For all $u$ and $v \in VG$, this pseudo-norm has the properties:

i. $\|v\| \geq G$, with equality if, and only if, $v = GGG$,

ii. $\|u + v\| \leq \|u\| + \|v\|$

There is a strong correlation between this pseudo-norm and interaction energy of amino acids in proteins (Table 5). Moreover, this pseudo-norm allows us interpret the genetic code as a non-dimensional code scale of interaction energies (see Sanchez et al. 2005b). Statistical evidence supporting this interpretation could be obtained from the linear regression analysis of codon pseudo-norm versus amino acid scales. The best fitting equation is:

$$\|X_1X_2X_3\| = 0.409 e_{ir1985} - 0.669 \Delta G_{water\text{-}oct} + 0.458 \Delta G_{CHP\text{-}water} - 0.406 Hydc + 0.888 \Delta G_{transfer} - 0.079 Polarity$$

These regression analyses are noticeable statistical evidences –highly significant– of the strong connection between the codon distance and the amino acid scales. The statistical summary of these regressions are presented in Table 6. In addition, the distance between wild type and mutant codons approach smaller values in mutational variants of four genes, i.e, human phenylalanine hydroxylase, human β-globin, HIV-1 protease and HIV-1 reverse transcriptase (Table 7).

These results suggest the usefulness of base order {A, C, G, U} to define the vector space of genetic code and the $LG$. The eight base orders mentioned above lead us to the same partition of the standard genetic code table; however, in practice it could be convenient to use that base order with the best physicochemical connections.

# 4. Conclusion

The established order in the Boolean lattice of the 4 bases of DNA allows define a new vector spaces and Lie algebra that help us to have a better understanding of the logic underlying the genetic code. Algebraic relationships among codons allow arrangements similar to previous models, obtained by intuition in the past century (Epstein, 1966; Crick, 1968). Our results suggest that algebraic relationships, codon assignment and physicochemical properties of amino acids are not connected at random and therefore the genetic code origin was not at random but certainly it follows deterministic rules.

Besides, this model could help understanding mutational events in the molecular evolution processes. Particularly, the distance defined in the vector space of genetic code was demonstrated to be strongly connected to interaction energy of amino acids in proteins. Moreover, the distance between wild type and mutant codons in genes tend to be the smallest values.

# References

Alf-Steinberger, C. (1969). The genetic code and error transmission. Proc. Natl. Acad. Sci. USA, 64, 584-591

Bashford, J.D., Jarvis P.D.: The genetic code as a periodic table. Biosystems **57**, 147-61 (2000)

Bashford, J.D., Tsohantjis, I., Jarvis, P.D.: 1998. A supersymmetric model for the evolution of the genetic code. Proc. Natl. Acad. Sci. USA 95, 987–992 (1998)

Beland, P., Allen, T.F.: The origin and evolution of the genetic code. J Theor Biol. 170, 359-365 (1994)

Birkhoff, G., MacLane, S.: A survey of Modern Algebra. The Macmillan Company. New York (1941)

Chothia, C.H.: Hydrophobic bonding and accessible surface area in proteins. Nature 248, 338-339 (1974)

Chothia, C.H.: Structural Invariants in Protein Folding. Nature 354, 304-308 (1975)

Crick, F.H.C.: The origin of the genetic code. J. Mol. Biol. 38, 367-379 (1968)

Eck, R.V.: Genetic Code - Emergence Of A Symmetrical Pattern. Science 140, 477-481 (1963)

Epstein, C.J.: Role of the amino-acid "code" and of selection for conformation in the evolution of proteins. Nature 210, 25-28 (1966)

Fauchere, J.L., Pliska, V.: Hydrophobic parameters pi of amino acid side chains from the partitioning of N-acetyl-amino-acid amides. Eur. J. Med. Chem. 18, 369-375 (1983)

Friedman, S.M., Weinstein, I..B. (1964). Lack of fidelity in the translation of ribopolynucleotides. Proc Natl Acad Sci USA 52, 988-996

Gillis, D., Massar, S., Cerf, N.J., Rooman, M.: Optimality of the genetic code with respect to protein stability and amino acid frequencies. Genome Biology 2, research0049.1–research0049.12 (2001)

Grantham, R.: Amino Acid Difference Formula to Help Explain Protein Evolution. Science 185, 862-864 (1974)

Jiménez-Montaño, M.A., 1996. The hypercube structure of the genetic code explains conservative and non-conservative amino acid substitutions in vivo and in vitro. Biosystems 39, 117-125.

Jukes, T. H. The amino acid code. In Comprehensive Biochemistry. Edited by. A. Neuberger, pp. 235 - 293. Amsterdam: Elsevier (1977)

Kostrikin, A. I.. (1980). Introducción al álgebra. Editorial MIR, Moscú

Lehmann, J., 2000. Physico-chemical Constraints Connected with the Coding Properties of the Genetic System. J. Theor. Biol. 202, 129-144.

Lewin, B. (2004). Genes VIII. Oxford University Press.

Parker, J. (1989). Errors and alternatives in reading the universal genetic code. Microbiol Rev. 53, 273-298

Pontriaguin, L.S.: Grupos Continuos. Capítulo 10. Editorial Mir. Moscow, pp. 338-451 (1978)

Redéi, L.: Algebra, Vol. 1. Akadémiai Kiadó, Budapest (1967)

Robin, D., Knight, R.D., Freeland. S.J., Landweber. LF.: Selection, history and chemistry: the three faces of the genetic code. Trends Biochem Sci. 24: 241-247 (1999)

Rose, G.D., Geselowitz, A.R, Lesser, G.J., Lee, R.H., Zehfus, M.H.: Hydrophobicity of amino acid residues in globular proteins. Sciences 229, 834-838 (1985)

Sánchez R., Morgado E., Grau R. (2005a). A genetic code boolean structure I. The meaning of boolean deductions, Bull. .Math. Biol., 67, 1–14.

Sánchez R., Morgado E., Grau R. (2005b). Gene algebra from a genetic code algebraic structure. J. Math. Biol. DOI: 10.1007/s00285-005-0332-8

Sánchez, R., Grau, R. and Morgado, E. (2004a). The Genetic Code Boolean Lattice. MATCH Commun. Math. Comput. Chem 52, 29-46

Sánchez, R., Grau, R., Morgado, E. (2004b). Genetic Code Boolean Algebras, WSEAS Transactions on Biology and Biomedicine 1, 190-197

Siemion, I.Z., Siemion, P.J., Krajewski, K.: Chou-Fasman conformational amino acid parameters and the genetic code. Biosystems. 36, 231-238.

Volkenshtein, M.V.: Biofísica. Editorial MIR, Moscú, Capítulo 17, 621-639 (1985)

Woese, C.R. (1965). On the evolution of the genetic code. Proc Natl Acad Sci USA, 54, 1546-1552

Woese, C.R.: Order in the genetic code. Proc. Natl Acad. Sci. USA 54, 71-75 (1965)

Yang, Z.: Adaptive molecular evolution. In Handbook of statistical genetics, (Balding, M., Bishop, M. & Cannings, C., eds), Wiley:London, pp. 327-50 (2000)

Zamyatin, A.A.: Protein Volume in Solution, Prog. Biophys. Mol. Biol. 24, 107-123 (1972)

# Tables

**Table 1.** Operation tables of the Galois field on the ordered set of 4 DNA bases {G, U, A, C}. Two analogous operation tables can be obtained for the dual set.

| Sum | | | | | Product | | | | |
|---|---|---|---|---|---|---|---|---|---|
| + | G | U | A | C | • | G | U | A | C |
| G | G | U | A | C | G | G | G | G | G |
| U | U | G | C | A | U | G | U | A | C |
| A | A | C | G | U | A | G | A | C | U |
| C | C | A | U | G | C | G | C | U | A |

**Table 2.** The standard genetic code table.

| | | Second base position | | | | | | | | |
|---|---|---|---|---|---|---|---|---|---|---|
| | | **U** | | **C** | | **A** | | **G** | | |
| First base position | **U** | UUU | ¹P | UCU | S | UAU | Y | UGU | C | U |
| | | UUC | | UCC | | UAC | | UGC | | C |
| | | UUA | L | UCA | | UAA | Stop | UGA | Stop | A |
| | | UUG | | UCG | | UAG | | UGG | W | G |
| | **C** | CUU | L | CCU | P | CAU | H | CGU | R | U |
| | | CUC | | CCC | | CAC | | CGC | | C |
| | | CUA | | CCA | | CAA | Q | CGA | | A |
| | | CUG | | CCG | | CAG | | CGG | | G |
| | **A** | AUU | I | ACU | T | AAU | N | AGU | S | U |
| | | AUC | | ACC | | AAC | | AGC | | C |
| | | AUA | | ACA | | AAA | K | AGA | R | A |
| | | AUG | M | ACG | | AAG | | AGG | | G |
| | **G** | GUU | V | GCU | A | GAU | D | GGU | G | U |
| | | GUC | | GCC | | GAC | | GGC | | C |
| | | GUA | | GCA | | GAA | E | GGA | | A |
| | | GUG | | GCG | | GAG | | GGG | | G |

¹The one letter symbol of amino acids.

**Table 3.** The elements of subsubgroup VG/C$_{GGA}$.

| GGR | G | GUR | V | GAR | E | GCR | A |
|---|---|---|---|---|---|---|---|
| GGN | G | GUY | V | GAY | D | GCY | A |
| TGR | W, - | UUR | L | TAR | - | TCR | S |
| TGY | C | UUY | F | TAY | Y | TCY | S |
| AGR | R | AUR | M, I | AAR | K | ACR | T |
| AGY | S | AUY | I | AAY | N | CAY | T |
| CGR | R | CUR | L | CAR | Q | CCR | P |
| CGY | R | CUY | L | CAY | H | CCY | P |

**Table 4**. Subsets of parallel codons in *LG*. If [X, Y] = GGG, then [X, X + Y] = GGG.

| X | | $^1P_X$ | Y | | $^1P_Y$ | $^2P_{mean}$ | X + Y | | $^1P_{X+Y}$ | $^3Max\Delta G_{w/o}$ | $^3Min\Delta G_{w/o}$ |
|---|---|---|---|---|---|---|---|---|---|---|---|
| $^4$V | GUG | 5.9 | E | GAG | 12.3 | 9.1 | A | GCG | 8.1 | 2.53 | 1.24 |
| V | GUU | 5.9 | E | GAA | 12.3 | 9.1 | A | GCC | 8.1 | 2.53 | 1.24 |
| V | GUA | 5.9 | D | GAC | 13 | 9.45 | A | GCU | 8.1 | 2.71 | 1.24 |
| V | GUC | 5.9 | D | GAU | 13 | 9.45 | A | GCA | 8.1 | 2.71 | 1.24 |
| L | UUG | 4.9 | K | AAG | 11.3 | 8.1 | P | CCG | 8 | 3.71 | 1.33 |
| F | UUU | 5.2 | K | AAA | 11.3 | 8.25 | P | CCC | 8 | 3.83 | 1.45 |
| L | UUA | 4.9 | N | AAC | 11.6 | 8.25 | P | CCU | 8 | 3.13 | 1.33 |
| F | UUC | 5.2 | N | AAU | 11.6 | 8.4 | P | CCA | 8 | 3.25 | 1.45 |
| M | AUG | 5.7 | Q | CAC | 10.5 | 8.1 | S | UCG | 9.2 | 1.97 | 0.25 |
| I | AUU | 5.2 | Q | CAA | 10.5 | 7.85 | S | UCC | 9.2 | 2.75 | 0.25 |
| I | AUA | 5.2 | H | CAC | 10.4 | 7.8 | S | UCU | 9.2 | 2.5 | 0.23 |
| I | AUC | 5.2 | H | CAU | 10.4 | 7.8 | S | UCA | 9.2 | 2.5 | 0.23 |
| L | CUG | 4.9 | - | UAG | | | T | ACG | 8.6 | 1.96 | 1.96 |
| L | CUU | 4.9 | - | UAA | | | T | ACC | 8.6 | 1.96 | 1.96 |
| L | CUA | 4.9 | Y | UAC | 6.2 | 5.55 | T | ACU | 8.6 | 1.96 | 0.96 |
| L | CUC | 4.9 | Y | UAU | 6.2 | 5.55 | T | ACA | 8.6 | 1.96 | 0.96 |
| X | | | Y | | | | X Å Y | | | | |
| G | GGG | 9 | G | GGG | 9 | 9 | G | GGG | 9 | 0 | 0 |
| G | GGT | 9 | G | GGA | 9 | 9 | G | GGC | 9 | 0 | 0 |
| G | GGA | 9 | G | GGU | 9 | 9 | G | GGC | 9 | 0 | 0 |
| G | GGC | 9 | G | GGU | 9 | 9 | G | GGA | 9 | 0 | 0 |
| W | UGG | 5.4 | R | AGG | 10.5 | 7.95 | R | CGG | 10.5 | 4.43 | 0 |
| C | UGU | 5.5 | R | AGA | 10.5 | 8 | R | CGC | 10.5 | 3.46 | 0 |
| - | UGA | | S | AGC | 9.2 | 9.2 | R | CGU | 10.5 | 1.32 | 1.32 |
| C | UGC | 5.5 | S | AGU | 9.2 | 7.35 | R | CGA | 10.5 | 3.46 | 1.32 |

[1] $P_X$, $P_Y$, $P_{X+Y}$: amino acid polarity according to Grantham polarity scale (1974).
[2] $P_{mean}$: $P_X+P_Y/2$.
[3] In last two columns, maximum ($Max\Delta G_{w/o}$) and minimum ($Min\Delta G_{w/o}$) of the absolute differences in the values of Δ-free energy of transfer from water to octanol (Fauchere and Pliska, 1983).
[4] One-letter symbol for amino acids.

**Table 5.** Correlation between contact energy and the mean of amino acid pseudo-norms.

| [1]Scale | Pearson coefficient | Statistical Signification |
|---|---|---|
| $e_{i(1985)}$ | 0.858453 | 0.000 |
| $e_{ir(1985)}$ | 0.847810 | 0.000 |
| $e_{i(1996)}$ | 0.844178 | 0.000 |
| $e_{ir(1996)}$ | 0.829829 | 0.000 |

[1]The contact energy $e_i$ and $e_{ir}$ were expressed in kcal/mol by way of the expression: $-0,6*q_i*e_i/2$ and $-0,6*q_i*e_{ir}/2$ [see 33, 34].

**Table 6.** Statistical summary of the regression analysis codon norm versus amino acid scales. The adjusted R Square is 0.952 and the Durbin-Watson statistic is 2.265.

| Lineal regression through the origin | Unstandardized Coefficients | | t | Sig. | 95% Confidence Interval for B | |
|---|---|---|---|---|---|---|
| Independent variables | B | Std. Error | | | Lower Bound | Upper Bound |
| $e_{ir}$ 1985 | 0,409 | 0,041 | 9,856 | 0,000 | 0,3256 | 0,4918 |
| [1] $\Delta G_{\text{water-oct}}$ | -0,669 | 0,121 | -5,55 | 0,000 | -0,9105 | -0,4274 |
| [2] $\Delta G_{\text{CHP-water}}$ | 0,458 | 0,072 | 6,386 | 0,000 | 0,3142 | 0,6015 |
| [3] $Hyd_C$ | -0,406 | 0,158 | -2,565 | 0,013 | -0,7230 | -0,0887 |
| [4] $\Delta G_{\text{transfer}}$ | 0,888 | 0,157 | 5,647 | 0,000 | 0,5726 | 1,2027 |
| [5] $Polarity$ | -0,079 | 0,017 | -4,619 | 0,000 | -0,1139 | -0,0450 |

[1] transfer free energy from water to octanol..
[2] transfer free energy from CHP to water.
[3] Consensus normalized hydrophobicity scale.
[4] transfer free energy.
[5] Grantham polarity scale (1974).

**Table 7**. Distance between the wild type and mutant codons in two human genes and two HIV-1 genes.

| [1] Human Beta Globin | | | [2] Human PHA | | | [3] Protease | | | [3] Reverse transcriptase | | |
|---|---|---|---|---|---|---|---|---|---|---|---|
| Wild Type | Mutant | Norm | Wild Type | Mutant | Norm | Wild Type | Mutant | Norm | Wild Type | Mutant | Norm |
| CCU | CAU | 0.5 | UAU | UGU | 1. | GCU | AUU | 1.25 | GCC | GUC | 1. |
| ACC | AUC | 1. | GCC | GAC | 0.5 | GCU | CUC | 1.4375 | GCA | GGA | 1.5 |
| GUG | GAG | 1.5 | GCC | CCC | 0.375 | GCU | ACU | 0.25 | GAC | GCC | 0.5 |
| GUG | AUG | 0.25 | GCU | GUU | 1. | GCU | GUU | 1. | GAC | GAG | 0.09375 |
| GUG | CUG | 0.375 | GCC | ACC | 0.25 | GAU | AAU | 0.25 | GAC | GAG | 0.09375 |
| GUC | UUC | 0.125 | GCC | GUC | 1. | GAU | GAA | 0.09375 | GAC | GGC | 1. |
| CAC | CAA | 0.03125 | GCC | UCC | 0.125 | GGG | GAG | 1. | GAC | AAC | 0.25 |
| GUC | UUC | 0.125 | GCC | GUC | 1. | GGG | GUG | 0.5 | GAG | GCG | 0.5 |
| GAA | CAA | 0.375 | GCC | GAC | 0.5 | GGU | AGU | 0.25 | GAG | AAG | 0.25 |
| CUG | CCG | 1. | GCC | GUC | 1. | GGU | AGU | 0.25 | GAA | GCA | 0.5 |
| GCU | GUU | 1. | CGA | ACA | 1.625 | CAU | UAU | 0.25 | GAA | GAC | 0.03125 |
| CAC | CAG | 0.09375 | GCA | GUA | 1. | AUA | GUA | 0.25 | GAA | GGA | 1. |
| GAU | GAA | 0.09375 | GCC | GGC | 1.5 | AUU | CUU | 0.125 | GAA | GGA | 1. |
| GAU | AAU | 0.25 | GCC | ACC | 0.25 | AUC | CUC | 0.125 | UUU | UAU | 1.5 |
| AAU | UAU | 0.375 | GCA | CCA | 0.375 | AUU | AUG | 0.03125 | UUC | CUC | 0.25 |
| GUC | GAC | 1.5 | GCA | ACA | 0.25 | AUC | ACC | 1. | GGG | GAG | 1. |
| GAA | AAA | 0.25 | GCU | UCU | 0.125 | AUC | GUC | 0.25 | GGA | GCA | 1.5 |
| GCC | GUC | 1. | GCU | ACU | 0.25 | AUC | ACC | 1. | GGA | GAA | 1. |
| AAG | GAG | 0.25 | GCC | ACC | 0.25 | AUA | GCA | 1.25 | GGA | CAA | 1.375 |
| GGC | GAC | 1. | GCC | GGC | 1.5 | AUA | GUA | 0.25 | GGA | UCA | 1.625 |
| GAU | AAU | 0.25 | GCC | CCC | 0.375 | AAG | AUG | 1.5 | GGA | ACA | 1.75 |
| GGU | CGU | 0.375 | GCU | GUU | 1. | AAG | AGG | 1. | GGA | GUA | 0.5 |
| GUC | CUC | 0.375 | GCC | GAC | 0.5 | AAA | AUA | 1.5 | GGA | GUA | 0.5 |
| GGC | GAC | 1. | GCA | GAA | 0.5 | AAA | AGA | 1. | GGA | GUA | 0.5 |
| CAC | UAC | 0.25 | GCA | GUA | 1. | CUC | UUC | 0.25 | CAU | UAU | 0.25 |
| GAG | AAG | 0.25 | UGC | UGU | 0.0625 | CUC | AUC | 0.125 | AUA | AUG | 0.0625 |
| AAC | AUC | 1.5 | UGU | GGU | 0.125 | CUC | CGC | 0.5 | AUA | ACA | 1. |
| CAC | CCC | 0.5 | UGU | CGU | 0.25 | CUC | GUC | 0.375 | AAA | CAA | 0.125 |

| CAC | UAC | 0.25 | UGC | UAC | 1. | CUC | UAC | 1.75 | AAA | AGA | 1. |
| UGU | UGG | 0.03125 | UGC | UCC | 1.5 | CUA | AUA | 0.125 | AAA | ACA | 0.5 |
| GCC | GUC | 1. | UGC | GGC | 0.125 | UUA | AUA | 0.375 | AAA | GAA | 0.25 |

[1] Human Phenylalanine Hydroxylase (PAH) variants (http://www.pahdb.mcgill.ca/), Human β-Globin variants (http://globin.cse.psu.edu/).
[2] Single point mutations of HIV-1 Protease and HIV-1 Reverse Transcriptase (http://resdb.lanl.gov/Resist_DB).

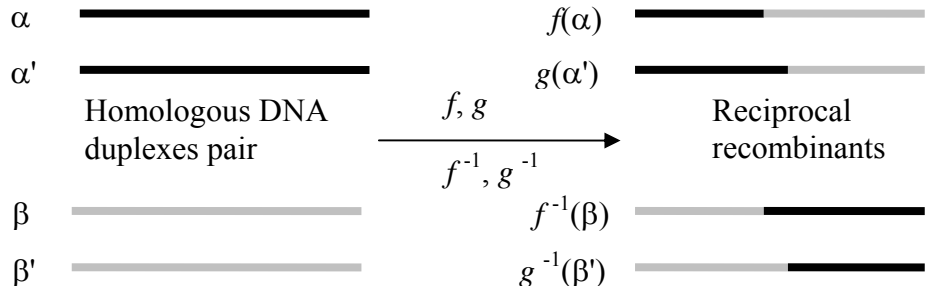

**Figure 1**. The homologous (generalized) recombination between two homologous DNA duplexes algebraically corresponds to the action of two automorphism pairs over two paired DNA duplexes. The symmetrical property of the commutator leads to the identities: $[\alpha, \beta] = [f(\alpha), f^{-1}(\beta)]$; $[\alpha', \beta'] = [g(\alpha'), g^{-1}(\beta')]$; $[\alpha, f(\alpha)] = [\beta, f^{-1}(\beta)]$; $[\alpha', g(\alpha')] = [\beta', g^{-1}(\beta')]$.